\documentclass[final,5p,times,twocolumn]{elsarticle}

\usepackage{graphics,graphicx,dcolumn,bm,fleqn,epic,eepic,float,ulem}
\usepackage{amssymb,amsmath,multirow,rotate,color}

\definecolor{red}{rgb}{1,0,0}
\definecolor{green}{rgb}{0,1,0}
\definecolor{blue}{rgb}{0,0,1}

\journal{Journal of Theoretical Biology}
\begin{document}

\begin{frontmatter}

  \title{Transition from endemic behavior
    to eradication of malaria \\ 
    due to combined drug therapies: an agent-model approach}
  
\author{Jo\~ao Sequeira$^{1,2}$}
\author{Jorge Lou\c{c}\~a$^1$}
\author{Ant\'onio M.~Mendes$^3$}
\author{Pedro G.~Lind$^{1,4}$}

\address{$^1$Instituto Universit\'ario de Lisboa (ISCTE-IUL), ISTAR-IUL,
  Av.~das For\c{c}as Armadas, 1649-026 Lisboa, Portugal}
\address{$^2$Hospital Santa Cruz, Av.~Prof.~Dr.~Reinaldo dos Santos,
  2790-134 Carnaxide, Portugal}
\address{$^3$Instituto de Medicina Molecular,
  Faculdade de Medicina, Universidade de Lisboa,
  Av.~Prof.~Egas Moniz, 1649-028 Lisboa, Portugal}
\address{$^4$Department of Computer Science,
  OsloMet -- Oslo Metropolitan University,
  P.O.~Box 4 St.~Olavs plass, N-0130 Oslo, Norway}


\begin{abstract}  
  We introduce an agent-based model describing a susceptible-infectious-susceptible (SIS) system of humans and mosquitoes to predict malaria epidemiological scenarios in realistic biological conditions.
  Emphasis is given to the transition from endemic behavior to eradication of malaria transmission induced by combined drug therapies acting on both the gametocytemia reduction and on the selective mosquito mortality during parasite development in the mosquito.
  Our mathematical framework enables to uncover the critical values of the parameters characterizing the effect of each drug therapy.
  Moreover, our results provide quantitative evidence of what is empirically known: interventions combining gametocytemia reduction through the use of gametocidal drugs, with the selective action of ivermectin during parasite development in the mosquito, may actively promote disease eradication in the long run.   
  In the agent model, the main properties of human-mosquito interactions are implemented as parameters and the model is validated by comparing simulations with real data of malaria incidence collected in the endemic malaria region of Chimoio in Mozambique.
  Finally, we discuss our findings in light of current drug administration strategies for malaria prevention, that may interfere with human-to-mosquito transmission process.
\end{abstract}

\begin{keyword}
Malaria spreading \sep
Agent models \sep
Gametocytemia \sep
Transmission control and mitigation
\end{keyword}

\end{frontmatter}


\section{Introduction}

Malaria is a parasitic disease, caused by a {\it Plasmodium} parasite, which
is still responsible for the death of nearly half a million individuals every
year worldwide \cite {WHO2018}. While some countries have had reasonable
success in rolling back malaria through well planned malaria prevention
interventions, malaria resurgence remains unpredictable.
Two types of factors may contribute to such unpredictability.
First, ''hidden'' factors, such as the asymptomatic
presence of gametocytes in human systemic circulation,
which are the precursors of male and female gametes of the parasite. 
Migration of just a few asymptomatic human, gametocyte carriers, into
particular African regions where the disease is controlled, may act
as a potential trigger of malaria outbreaks \cite{Lee2010, Pages2018}.
The presence of gametocytes may be mitigated through the application of
gametocidal drugs, such as primaquine or methylene blue.

Second, malaria transmission can be promoted due to the intrinsic
heterogeneity in human demography and mosquito behavior \cite{Lloyd1996}.
For example, in a potential outbreak
human fatality rate may rise out of proportion due to the
weaker immunity of local populations from reduced exposure to the
parasite \cite{Tyrrell2017}.
Or, in regions under anti-malaria massive drug administration,
drug-resistant parasite strains can develop and consequently, through
human migratory phenomena, they may be imported into areas of near eradication,
locally strengthening malaria transmission \cite{ Lee2010, Pages2018, DePina2018}.

Although the decisive role these factors can play, malaria transmission
mechanisms is well established, how their impact on malaria transmission
is modelled by the combined effect of different drug therapies in
heterogeneous populations is still not fully understood. 

The life cycle of {\it Plasmodium}, and its etiology may be summarized
as follows.
The malaria vector, the mosquito, \textit{Anopheles spp.}, usually lives, mates and feeds within a few miles distance from its birthplace \cite{Kaufmann2004}.
To become infectious to humans, the mosquito needs to survive 10 or more days
after feeding on a {\it Plasmodium} gametocyte carrier.
This time period is required to complete parasite sporogonic development inside the mosquito \cite{Eckhoff2011}, after which,
mosquito-to-human transmission becomes possible.
Therefore, stricly gametocidal drugs cannot only block human-mosquito transimission but can also have a strong impact on it.

Other drug agents, such as ivermectin, have become a promising antimalarial interventions due to its anophelocide properties and its ability in preventing parasite's maturation inside the mosquito \cite{Chaccour2010, Kobylinski2012,omura2017}.
It is known that mosquitoes, feeding on human hosts under ivermectin treatment, have a considerably lower life expectancy, with a large population of mosquitoes dying within 4 days after the blood meal \cite{Chaccour2010, Kobylinski2012, Ouedraogo2015}.
Moreover, interventions including mass administration of ivermectin in prevention of several other African endemic parasites resulted in a significant reduction in malaria endemic behavior on those regions \cite{Alout2014,mendes2017,kobylinski2011}.

To tackle the specific problem related with malaria transmission in a human community several mathematical models have been proposed.
Early models, such as those by Ross and Mcdonald, were deterministic \cite{Ross1915, Macdonald1957}, having nonetheless a significant relevance in malaria epidemiology \cite{Dietz1974, Koella1991, Ngwa2000,  Mandal2011, Chitnis2012, Chitnis2018, white2009} and being since then refined.
More recent variants have been developed with the help of modern satellite imaging, precise weather and geographical information,  computational agent-based modeling, and advanced statistics, such as hidden Markov processes, time-series analysis and big-data approaches \cite{ Eckhoff2011, Chitnis2012, Depinay2004, McKenzie2005, Gaudart2009, Kamgang2015, Ewing2016, Sarkar2017}.
In particular, agent-based models strengthen the importance of malaria simulation for disease prevention \cite{Eckhoff2011, Gerardin2015}. Based on the classical susceptible-infected (SI) model by Kermack and McKendrick,
stochastic modeling approaches were also proposed, with the aim of better implementing the uncertainties inherent to the disease dynamics \cite{Ferrao2017b}.

Epidemiological field data of malaria transmission is commonly presented as human monthly or weekly disease incidence \cite{Ferrao2017b, Aregawi2014}, while mosquito infection rates are obtained from data collected through the use of mosquito trapping devices \cite{Bomblies2009}.
However, since both are important to understand the transmission dynamics,
one should account for the combined effect of human and mosquito infection
prevalences.

In this paper, we introduce an agent-based model of malaria endemic/epidemic behavior, incorporating both human-to-mosquito and mosquito-to-human transmission processes. We parameterize some of the most important biological aspects of disease transmission, focusing mainly in the parameters describing the reduction of gametocytemia prevalence in the human host and the extension of ivermectin administration in the population.
The model assumes a typical isolated African village with limited access to drug therapy and is based on discrete Markov processes describing the succession of human-mosquito encounters, which are implemented through a Monte Carlo algorithm.
Tuning the parameter controlling the gametocytemia inside the human host
or the parameter controlling the fraction of the human population under
ivermectin treatment we uncover a phase transition between disease
eradication and epidemic prevalence.
In both cases, the transition is sensitive to minor changes in the
parameters and through mathematical analysis, we are able to predict
critical values separating the two phases, eradication and endemic prevalence.

We start in Section \ref{sec:model} by describing the agent model and the main parameters driving gametocytemia and human-mosquito infection dynamics. In Sections \ref{sec:results} and \ref{sec:validation}
we present respectively the main results and describe the validation
procedure using data sets from the endemic region of Chimoio in Mozambique.
In Section \ref{sec:discussion} we discuss the impact of our results on
possible clinical and medical strategies and conclude the paper.
\begin{table*}[t]
\begin{center}
\begin{tabular}{|l|c|c|}
\hline
\textsc{\large{Tunable parameters}} & \textsc{\large{Symbol}} & \textsc{\large{Value}}\\
\hline
Probability of ivermectin treatment & $p^{(iv)}$   &\rmfamily 0.00-0.10 \\
Duration of positive gametocytemia  & $\tau_{g}$ &\rmfamily 58-90 (days) \\
\hline
\hline
\textsc{\large{Fixed parameters}} & \textsc{\large{Symbol}} & \textsc{\large{Value}}\\
\hline
Number of human individuals    & $H$ & \rmfamily 2000\\
Number of (female) mosquitoes  & $M$ & \rmfamily 4000 \\
Average number of bites from one mosquito & $n_b$ & \rmfamily 2 (per day) \\
Total simulation time          & --  & \rmfamily 30 (years)\\
Maximum time of human infection (including time of parasite development) & $\tau_{d}$ &\rmfamily 150 (days) \\
Minimum time of human infection (including time of parasite development) & $\tau_0$ & \rmfamily 25 (days) \\
Average human infectious period, cf.~Eq.~(\ref{qh}) & $\tau_{c}$ & \rmfamily 87.5 (days) \\
Maximum life time of one mosquito & $\tau_{\rm max}$ &\rmfamily 40 (days)\\
Minimum life time of one mosquito & $\tau_{\rm min}$ &\rmfamily 0 (days)\\
Mosquito death probability
from feeding in human with ivermectin & $g_{iv}$ & \rmfamily 0.5 \\
Time needed to acquire immunity due to persistent reinfection & -- & \rmfamily 5 (years) \\
Time needed for losing immunity (in the absence of infection) & -- &\rmfamily 2 (years) \\
Probability of protection from LLIN, ITN and IRS barriers & $u$ & \rmfamily 0.25 \\
Single episode mosquito mortality of LLIN/ITN/IRS protection & $g_{irs}$ & \rmfamily 0.50 \\
Probability of a mosquito bite in the low season & $p_{ls}$ & \rmfamily 0.5 \\
Probability of a mosquito bite in the high season & $p_{hs}$ & \rmfamily 1 \\
Fraction of humans among all animals biten (anthropophilic factor) & $p_{Q}$ &\rmfamily 0.9 \\
Duration of the gonotrophic reproductive cycle & $\tau_{s}$ &\rmfamily 4 (days)\\
Duration of the high transmission season & $\delta_s$ &\rmfamily 150 (days)\\
Probability of human infection after infectious mosquito bite & $k_h$ &\rmfamily 0.20 \\
Probability of mosquito infection after bite in infectious human & $k_m$ &\rmfamily 0.20 \\
Time for parasite development in the mosquito & $\tau_{lm}$ &\rmfamily 10 (days)\\
Time for parasite development to gametocyte stage inside human host &  $\tau_{lh}$  & 10 (days) \\  
Probability of full protection due to acquired immunity & $v_{max}$ &\rmfamily 0.3 \\
Fraction of children (age $<$ 5 years) in the population & -- &\rmfamily 0.12 \\
Probability of positive gametocytemia in children & -- & \rmfamily 0.70 \\
Average number of humans that die (global causes) & $\mu_h$ & \rmfamily 0.015 (per year) \\
Seasonality overall bite probability & $\sigma$ & 0.7055\\
Initial protection probability from acquired immmunity & $\nu_0$ & 0.1 \\
Probability of mosquito bite from surviving mosquitoes past latency & $\pi_{lm}$ & 0.686 \\
\hline
\end{tabular}
\end{center}
\caption{Parameters of the agent-based model for malaria spreading within two
  interacting communities of human individuals and mosquitoes.
  The values chosen for the simulations are taken, based in previous
  studies. See Refs.~\cite{Eckhoff2011,Macdonald1957,Mandal2011,Chitnis2012,Gaudart2009,Ferrao2017b,Karl2011,Felger2012,Gurarie2012,WHO2013,Ngonghala2016,Filipe2007,Doolan2009,Bretscher2015,Coffeng2017}.}
\label{tab01}
\end{table*}
\begin{figure*}[t]
\centering
\includegraphics[width=0.85\textwidth]{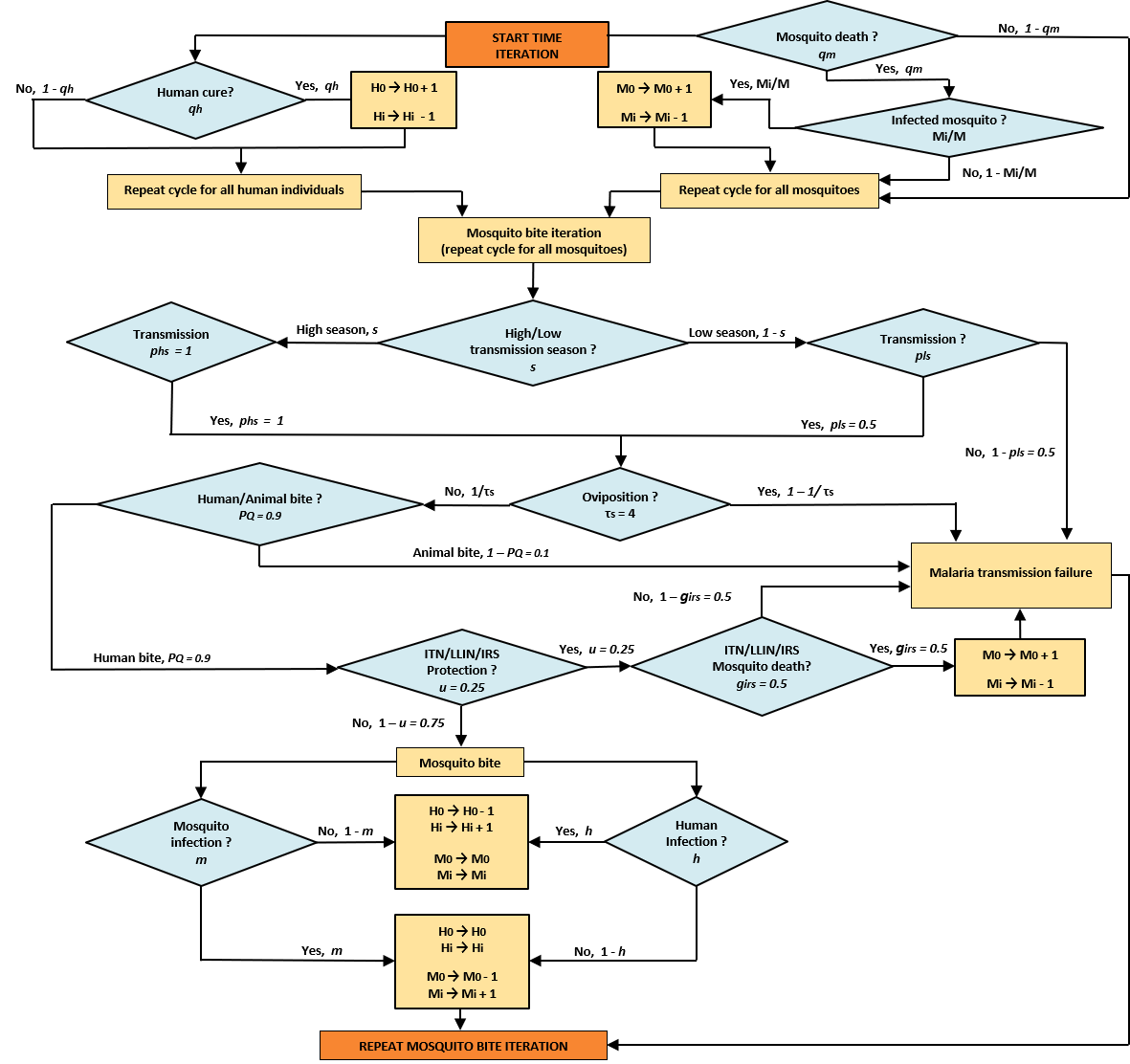}
\caption{\protect
  Flowchart of the agent-based model for human-mosquito interaction
  to reproduce scenarios of malaria spreading.
  Probabilities $q_h$ and $q_m$ are given in
  Eqs.~(\ref{qh}) and (\ref{qm}) respectively.
  The other probabilities are given in Tab.~\ref{tab01}.
  The probability for infecting a human or a mosquito depend
  on $p_h$ and $p_m$, given in Eqs.~(\ref{ph}) and (\ref{pm})
  respectively, and also on additional details concerning
  the dynamics of immunity acquisition of each human individual
  and the fraction of the human population composed by children
  (see text).}
\label{fig01}
\end{figure*}

\section{Agent model for human-to-mosquito and mosquito-to-human transmission}
\label{sec:model}

We consider a systems of $M=4000$ mosquitoes and $H=2000$ human
individuals, where each population is divided into a number of
healthy and another of infected individuals, represented by $H_0$ and $M_0$
and by $H_i$ and $M_i$ respectively: $H=H_0+H_i$ and $M=M_0+M_i$.
Although in reality, the density of mosquitoes is much higher, we
model the amount of mosquitoes as the effective fraction of
the overall mosquito mass, imposing an average of two bites per
day for each mosquito.

The chosen values of each parameter are given in Tab.~\ref{tab01},
and the algorithm keeps track of all attributes for each agent, human
or mosquito, in particular age, time since infection, and immunity
status.
Notice that only two parameters are modulated, namely the fraction
$p_{(iv)}$ of the human population subjected to ivermectin treatment
and the effectiveness of gametocidal drugs, measured as the number
$\tau_g$ of days of positive gametocytemia.
All other parameters are kept at constant values and their values
were chosen according to previous studies.

The flowchart describing the computer implementation of the agent model
is sketched in Fig.~\ref{fig01} and is described as follows.
The algorithm simulates a total time interval of $30$ years and it
starts by evaluating each individual, to ascertain if
it became cured or not.
In the case of human individuals the recovered rate $q_h$ is a fixed
value, dependent on the average
time $\tau_c$ it takes for one individual to be cured,
\begin{equation}
  q_h = \frac{1}{\tau_c} \, .
  \label{qh}
\end{equation}
In the case of mosquitoes, there is no explicit recovery rate.
Every dead mosquito is replaced by a new healthy mosquito. As such,
the mosquito recovery rate equals its mortality rate
$q_m$. The mosquitoes' mortality is determined by its natural
life expectancy, $\tau_m$, the fraction $p^{(iv)}$ of human with whom the
mosquito interacts that is under ivermectin treatment and the life
expectancy $\tau_m^{(iv)}$ of a mosquito with exposure to ivermectin:
\begin{equation}
  q_{m} = (1-p^{(iv)})\frac{1}{\tau_m}+p^{(iv)}\frac{1}{\tau_m^{(iv)}} \, .
  \label{qm}
\end{equation}

The two rates, $q_h$ and $q_m$, are not directly implemented in the agent
model.
Instead, we impose a maximum time of human infection
of $\tau_{d}=150$ days and a minimum time of 25 days, uniformly distributed,
yielding an average human infectious period of $\tau_{c}=87.5$ days,
a maximum mosquito life time of 40 days and a minimum
life time of 0 days, uniformly distributed,
yielding an average life expectancy of one mosquito $\tau_{m}=10$ days, as well
as a probability $g_{iv}=0.5$ of one mosquito to die from feeding in human
host under ivermectin treatment.
In case of an infectious mosquito bite in an infected human host, a human
reinfection or super-infection
occurs\footnote{Persistent reinfection is defined as
  a new contact between an infected human host and an infected
  mosquito, during the
  time period of active infection. In practice, since the
  average time of infection in one human is 87.5 days,
  one human host may reacquire a new malaria infection within three months
  after the initial infection episode, thus perpetuating disease
  transmission as well as immunity individual acquisition.} 
and the disease time of that human
individual is reset to half of the present disease time.
Beyond 5 years of persistent human reinfections, the human host acquires
maximum immunity and after 2 years with no infection events, the host loses
immunity completely.

In case the mosquito succeds in overcoming the barrier protection,
the algorithm starts to ascertain if transmission will take place
or not. This is done by computing the probability $r$ for one mosquito
and one human individual or other animal to contact through one bite, which is
given by
\begin{equation}
  r=\Big ( (1-s)p_{ls}+sp_{hs} \Big )\frac{p_{Q}}{\tau_s}\, ,
\label{r}
\end{equation}
where $s$ is the fraction of time in the year with high disease transmission
(percentage of time in rainy season),
$p_{ls}$ and $p_{hs}$ represent the fractions of the year covered by the low and
high seasons respectively,
$p_Q$ is the fraction of humans among all animals able to be
biten by  one mosquito within the geographical region covered
by the mosquito community, and 
$\tau_s$ is the duration of the gonotrophic reproductive cycle.
In the agent model, we use values provided in previous studies,
namely $p_{ls}=0.5$, $p_{hs}=1$, $p_{Q}=0.9$ and $\tau_{s}=4$ days.
Moreover, inspired in Mozambique seasonality\cite{Ferrao2017b, Ferrao2017a},
we consider 150 days for the duration of the high transmission season,
i.e.~$s=150/365$.
Notice that during transmission season, one considers a non-zero probability
of transmission; in this way, disease transmission may occur during the whole
year, although with higher intensity during the high-transmission season.

Upon updating the number of healthy humans individuals and mosquitoes,
the algorithm proceeds to generate one mosquito bite attempt.
Here one introduces the probability $u=0.25$ that long lasting
insecticide-impregnated nets (LLIN), insecticide-impregnated
nets (ITN) or indoor residual spraying (IRS) may
protect human hosts from mosquito bites.
This parameter represents the degree of human population protection
resulting from LLIN, ITN or IRS preventive measures, and simulates
the probability of mosquito bite failure due to protective barrier.
Additionally, we also introduce the effect of barriers
in killing the mosquito during the bite attempt.
In the model, the probability of mosquito mortality induced by
protective barriers is 0.5.
\begin{figure*}[t]
\centering
\includegraphics[width=0.95\textwidth]{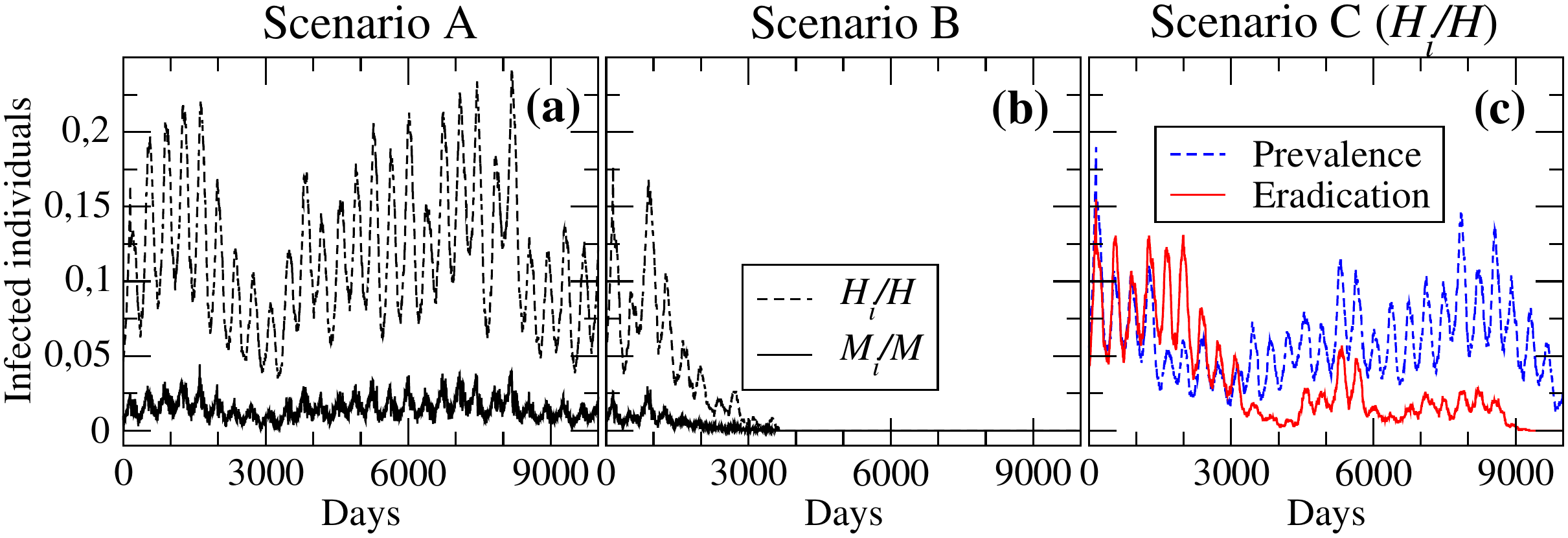}
\caption{\protect
  Illustration of the three scenarios tuned by gametocytemia
  parameter $w_h$:
  {\bf (a)} Scenario A, disease epidemic prevalence ($w_h=0.453$), 
  {\bf (b)} Scenario B, disease eradication ($w_h=0.387$), and
  {\bf (c)} Scenario C, transition between prevalence
  and eradication ($w_h=0.420$). In all cases $p_{iv}=0$.}
\label{fig02}
\end{figure*}

In the case one of the above factors succeeds, malaria transmission fails.
In case all barriers fail, the algorithm finds one mosquito-human interaction
through one bite.
If both interacting human hosts are infected or
both are not infected, both populations remain unchanged and the algorithm
starts the next iteration.
If only one individual (either human or mosquito) is infected, the algorithm
ascertains if malaria transmission is successful.

The probability $p_h$ for such single bite to effectively transmit the
parasite from an infected mosquito to a healthy human depends on three
factors, namely (i) the fraction $M_i/M$ of infected mosquitoes, 
the probability $k_h$ to get infected from one single mosquito bite\footnote{The value of the probability to get infected from one single bite, either for humans as for mosquitoes, is given by the inverse of the number of infected mosquito bites necessary to infect one human or one mosquito and it estimated from controlled malaria infection, in laboratory settings.}
and the probability $w_m$ that the
mosquito is ready to transmit the parasite, yielding
\begin{equation}
  p_h=\frac{M_i}{M}k_hw_m\, ,
\label{ph}
\end{equation}
and similarly, the probability for one single bite to effectively transmit
the parasite from an infected human to a healthy mosquito is
\begin{equation}
  p_m=\frac{H_i}{H}k_hw_h\, .
\label{pm}
\end{equation}

The probability $w_m$ is obtained from the fraction of
surviving mosquitoes past the period of parasite development,
i.e.
\begin{equation}
  w_{m} = e^{-q_m\tau_{l}} \, ,
  \label{wm}
\end{equation}
where the parameter $\tau_l$ is the period of parasite development
in the mosquito.
As for $w_{h}$ it measures the fraction of time of the duration
of positive gametocytemia from the maximal period of human infection,
\begin{equation}
  w_{h} = \frac{\tau_{g}}{\tau_{d}} \, ,
  \label{wh}
\end{equation}
where $\tau_{d}$ is the maximal period of human infection
and $\tau_{g}$ is the duration of positive
gametocytemia.

In the agent model we fix $k_m=k_h=0.2$, $\tau_{lm}=\tau_{lh}=10$ days and
$\tau_d=150$ days.
Notice that the duration of positive gametocytemia is a
tunable parameter used for varying $w_h$, which will be one of
the important parameters below.
Since $\tau_d$ takes values between 58 and 90 days (see Tab.~\ref{tab01}),
the probability $w_h$ varies between 0.387 and 0.733\footnote{%
Gametocyte detection threshold by light microscopy usually retrieves measurements between 5 and 10 gametocytes per $\mu$L. But with current molecular detection methods, that threshold may be as low as 0.1 per $\mu$L \cite{Karl2011}. 
It is assumed that during the period of human disease, gametocytemia will occur according to a random stochastic process, with a predefined probability of human-to-mosquito transmission at every mosquito bite in the range of admissible values \cite{Karl2011,Kuehn2010}.}, a range that
includes a phase transition from malaria eradication to malaria endemic
behavior.

Notice that a higher gametocyte density will result in higher human-to-mosquito transmission efficiency.
Consequently, the concept of gametocytemia reduction is considered equivalent
to the effects of treatment with gametocidal agents such as primaquine or
methylene blue.

The agent model implements three additional ingredients that are not usually
taken into account in simulation of malaria transmission dynamics.

First, 
in the present model we simulate the use of ivermectin in a fraction of the
human population ($p_{iv}$), assuming a global ivermectin-related mosquito fatality rate ($g_{iv}$) of 0.5.

Second, to consider the effect of acquired immunity to malaria infection.
Acquired immunity $v$ against malaria changes according to the history
of infection and the genetic traits of a particular human individual.
The value of $v$ can increase in the case of repeated reinfections,
or decrease, in case no infection is observed during a certain time.
The time to acquire protective immunity after every
infection episode is typically longer than that of the immunity loss.
We consider that if the human host does not contact with the parasite
during two years, he/she loses the acquired immunity against the parasite,
while maximal immunity is gained after 5 years with persistent reinfection.
Moreover, maximum protective immunity is different from complete protective
immunity, as a human can not be more
than 30\% immune, $v_{max}=0.3$.

Third, the extreme vulnerability to malaria infection of children under
5 years of age is a well known critical factor in the disease morbidity and
mortality.
We therefore consider additional effects for the subgroup of children
in the human population.
A simplified age effect is considered: the fraction of children under 5 years
is 12\% of the total human population and for all children under 5 years,
immunity is considered to be absent, with a higher gametocytemia
prevalence during disease duration, namely during 70\% of the
time \cite{Gurarie2012,Filipe2007, Doolan2009, Bretscher2015, Coffeng2017}.

Malaria unrelated human mortality, is also considered in our model. However, its magnitude is considered low, namely $0.015$ cases per year, i.e.~it has negligible effects in disease transmission.
The system is always initialized with a fraction of infected mosquitoes of 1\%,
a fraction of infected humans of 5\% and an initial acquired immunity of
$v_{0}=0.1$ for every human individual\footnote{Except children under 5 years, who are assumed to have an acquired immunity of 0.4.}.

\section{Assessing the effect of drug therapies}
\label{sec:results}

In this section, we address separately the effect of gametocidal drugs and
of ivermectin, choosing proper values for generating each of three
possible scenarios:
\begin{itemize}
\item Scenario A: Disease endemic/epidemic prevalence ($H_i>0$ and $M_i>0$).
\item Scenario B: Disease eradication ($H_i=0$ and $M_i=0$).
\item Scenario C: Critical phase transition between endemic disease and eradication, where some of the simulations evolve to disease eradication, while other to epidemic prevalence.
\end{itemize}
  
\subsection{The role of gametocytemia in disease dynamics}
\label{subsec:gametocytemia}

We define the effect of ivermectin as null at $p_{iv}=0$ 
and generate illustrative examples of each scenario.
For Scenario A, we consider $\tau_g=68$ days of gametocytemia yielding
a value  of $w_h = 68/150= 0.453$, for Scenario B we consider $\tau_g=58$ days
of gametocytemia, i.e.~$w_h = 0.387$, and for Scenario C $\tau_g=63$ days
($w_h=0.420$).
Results are shown in Fig.~\ref{fig02}.

Figure \ref{fig02}a illustrates Scenario A, where 
both human and mosquito communities evolve in periodic cycles,
reflecting the seasonal character of malaria incidence,
changing between low and high transmission seasons.
Here, none of the infected communities converges to eradication.
In Fig.~\ref{fig02}b one observes the opposite:
both communities eventually get cured with no cases of infection.
In the plotted example this occurs after one seasonal cycle (1 year).
For the Scenario A, we obtained $12\%\pm4\%$ of infected humans and
$1.5\%\pm 0.7\%$ of infected mosquitoes, while for the Scenario B,
we obtained $1\%\pm 3\%$ of infected humans and
$0.2\%\pm 0.4\%$ of infected mosquitoes.
\begin{figure}[t]
\centering
\includegraphics[width=0.4\textwidth]{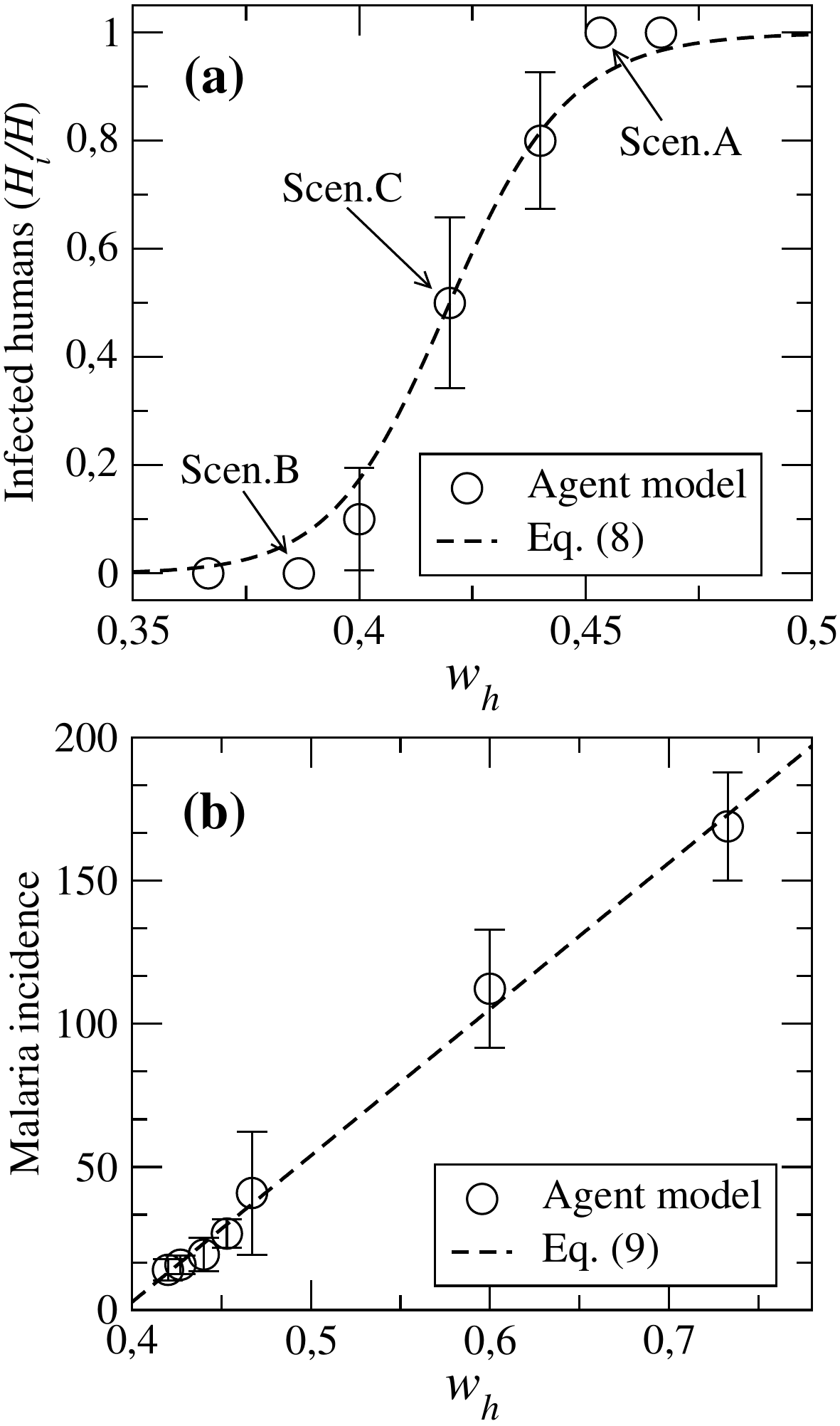}
\caption{\protect
  {\bf (a)}
  Probability of epidemic outcome with changing gametocytemia duration
  at phase transition. The three scenarios illustrated in Fig.~\ref{fig02}
  are indicated with arrows. The function in Eq.~(\ref{F_gam})
  is plotted with dashed line.
  The fraction of infected humans is averaged over $10$ realizations
  for each value of $w_h$.
  {\bf (b)}
  Annual malaria incidence per 100 habitants and its correlation with
  the positive gametocytemia $w_{h}$.
  Correlations were computed by averaging over $10$ realizations for
  each value of $w_h$.}
\label{fig03}
\end{figure}


In Fig.~\ref{fig02}c we observe the intermediate situation, between
endemic prevalence and eradication.
Two different outcomes occur at identical gametocytemia levels:
in black dashed lines we plot the evolution of human community in a
simulation where the disease persists for more than 30 years 
and in red solid lines, one observes the resulting community
evolution towards a state of eradication, after around 20 years.
This intermediate scenario occurs for $w_h\sim 0.42$.
For the eradication case of Scenario C we obtained
$3\%\pm 3\%$ of infected humans and $4\%\pm 4\%$ of infected mosquitos,
and for the prevalence situation,
$6\%\pm 3\%$ of infected humans and $0.7\%\pm 0.4\%$ of infected
mosquitos.


Two important features must be addressed at this point.
First, the time span needed for eradication at the transition value
is considerably larger than for values below the transition.
This is a common feature in critical phase transitions \cite{stanleybook}.

Second, the different outcomes from Scenarios A, B and C result from
small changes in the time of gametocytemia prevalence:
the differences between scenarios A, B and C are not greater than 5 days,
which represents gametocytemia differences of $\pm 3.3 \%$.
Consequently, small changes in gametocytemia status may result in
significant changes on the level of epidemic outcome, a feature that
shows the importance of gametocytemia in controlling malaria transmission. 

To better uncover the transition from endemic prevalence to eradication
due to gametocytemia control,
we generate $10$ different realizations for a set of different
$w_h$ values within a range covering all three scenarios.
Results are shown in Fig.~\ref{fig03}a.
As one sees,
while for Scenarios A and B, all realizations converged to the same
state, prevalence or eradication respectively, for Scenario C one fraction
of the realizations ended in endemic prevalence while the rest converged to
eradication.
Therefore, we argue that there is a critical value of
gametocytemia days that guarantees full recovery of the community.

A quantitative approach for estimating this transition gametocytemia
value is to approximate the transition curve in Fig.~\ref{fig03}a
by a step function of the form
\begin{equation}
  F_{g} (w_{h})= \frac{1}{1+ \left(\frac{w_{h}^{(t)}}{w_{h}}\right)^{\alpha_{g}}} \, ,
  \label{F_gam}
\end{equation}  
yielding an estimate for the transition gametocytemia value of
$w_{h}^{(t)}=0.42$ and for the exponent $\alpha_{g}=32$.

In real situations of malaria epidemics, there are several difficulties
in properly determining the annual malaria
incidence\footnote{Annual malaria incidence represents the instant expected
  average of malaria incidence per 100 inhabitants during one full year,
  if transmission conditions remains unchanged.},
which is an adequate measure for 
evaluating the gravity and extension of the epidemic.
Through simulations the annual malaria incidence can not only
be more easily calculated, but it is also possible to investigate
how it relates with other variables. As shown in Fig.~\ref{fig03}b,
we observe a clear linear relation between the average malaria
incidence $I$ and the gametocytemia parameter $w_{h}$.
In the plot, for each value of positive gametocytemia $w_h$
we obtained the average of the annual malaria incidence over 10 different
simulations.
A linear regression of the simulation results yields
\begin{equation}
    I = -202 + 508 w_{h} \, ,
    \label{Reg01}
\end{equation}
with a coefficient of determination of $r^{2} = 0.9983$ and a $p$-value of
$P<0.001$.
Notice that for $w_h<202/508\lesssim w_h^{(t)}$, close to obtained transition
value of gametocytemia, the annual incidence
is negative, meaning that the system converges to a scenario
of disease eradication. Only for values above the transition value
one observes a positive malaria incidence.

\subsection {The role of ivermectin in transmission prevention}
\label{subsec:ivermectin}

To investigate the role of ivermectin we fix value for the time of
positive gametocytemia, since it appears to be independent from the
efficiency in human-to-mosquito transmission efficiency.
We choose a stable epidemic background with 90 days of gametocytemia,
corresponding to $w_{h}=  0.6$.
To investigate the mosquito mortality due to ivermectin's
influence, we first focus on
three different values of the fraction of human population
under ivermectin treatment, namely $p_{iv}=0,0.05,0.1$.
Each of these three values illustrate one of three different regimes,
respectively (i) absence of ivermectin treatment,
(ii) weak ivermectin administration and
(iii) moderate ivermectin administration.

Our results show that, while in the absence of ivermectin administration
the mosquito mortality during parasite development is 79.6\%,
for $p_{iv}=0.05$ the mortality increases to 84.4\% and for
$p_{iv}=0.1$ to 88.1\%.
In the case of bite failure due to barrier protection, the mosquito
mortality is considered relevant, and set at
$g_{irs}=0.5$\footnote{There is no precise knowledge concerning
  the probability for the mosquito to die due to ITN, IRS or LLIN barriers.
  We assumed a value of 0.5, which together with a coverage of ITN of
  25\%, results in a global
  mortality due to ITN barriers of $0.25\times 0.5=0.125$.
  This value is probably below the real value,
  since in several African countries
  the LLIN/ITN/IRS coverage may be as high as 80\%.}.

We have also observed that, in the case of ivermectin random usage in 5\% of
the population, disease eradication may occur roughly 20 years later. But
if ivermectin is administered to 10\% of the human population,
a disease eradication outcome may be possible much earlier (less than 4 years).
Moreover, the administration of ivermectin induces a reduction in the
frequency of healthy mosquito bites in an infected human (not shown).
\begin{figure}[t]
\centering
\includegraphics[width=0.4\textwidth]{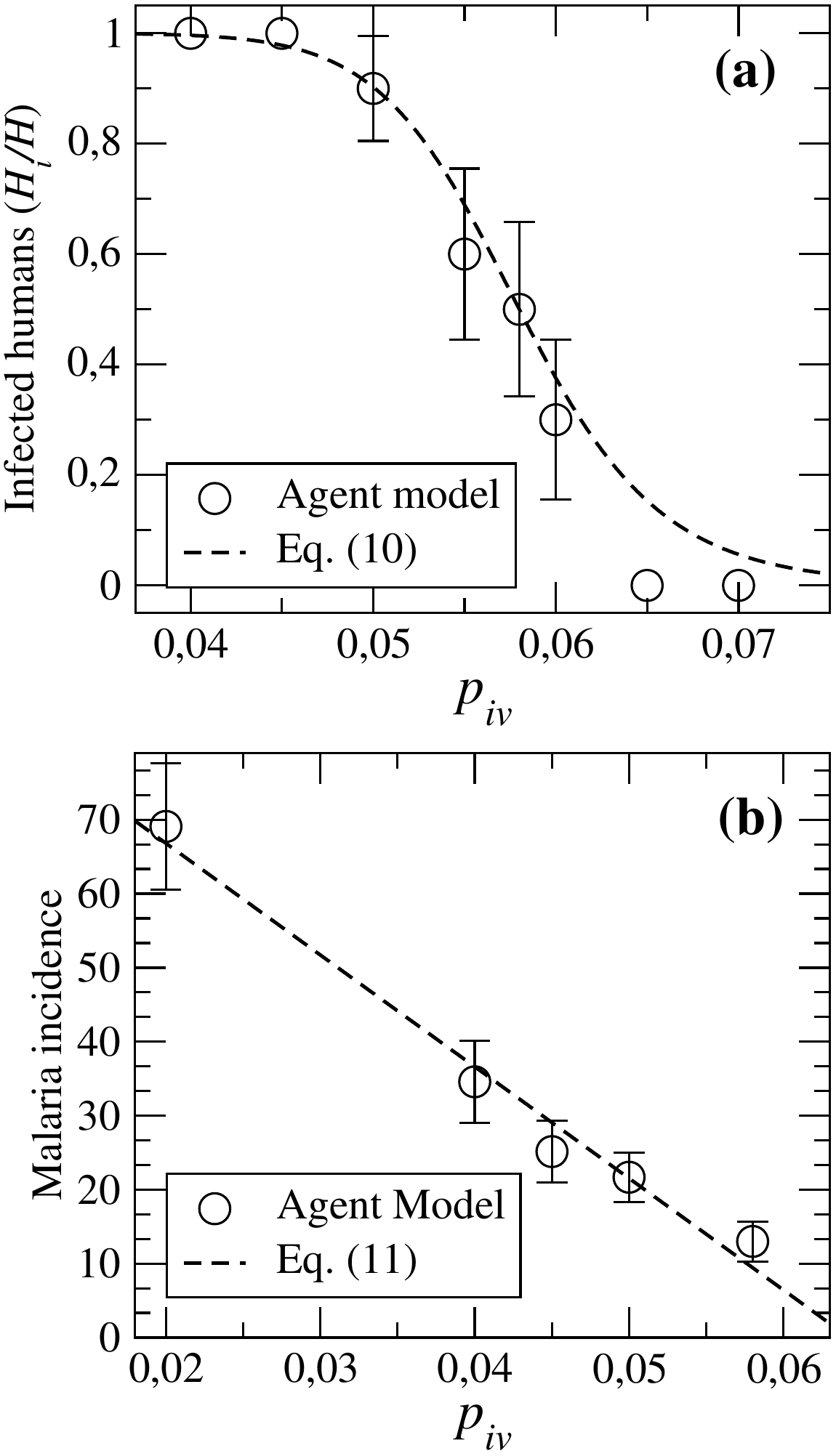}
\caption{\protect
  {\bf (a)}
  Probability of epidemic outcome with probability of ivermectin treatment 
  at phase transition, and approximate fit function.
  Here we run $10$ trials for each value of $p_{iv}$.
  {\bf (b)}
  Annual malaria incidence per 100 habitants and correlation with
  ivermectin treatment probability ($p_{iv}$).
  Positive gametocytemia is fixed at $w_{h}=0.6$.}
\label{fig04}
\end{figure}

Varying the fraction $p_{iv}$ uncovers also a continuous phase transition
from prevalence to eradication. See Fig.~\ref{fig04}a. Differently from
the transition by gametocytemia variation, here the phase transition
is only visible for high gametocytemia levels, typically $w_{h}=0.6$
or larger. A possible explanation for this is the strong inhibitory
effect of ivermectin on human-to-mosquito disease transmission.

In Fig.~\ref{fig04}a, a phase transition from epidemic
prevalence to disease eradication can be observed.
Here, the critical value of the human fraction with ivermectin is
approximately $p_{iv}=0.058$.
Higher $p_{iv}$ values induce faster disease eradication scenarios.
Similarly to Eq.~(\ref{F_gam}), the step function can be modelled
through the function
\begin{equation}
  F_{iv} (p_{iv})= 1 - \frac{1}{1+ \left(\frac{p_{iv}^{(t)}}{p_{iv}}\right)^{\alpha_{iv}}} \, .
  \label{F_iv}
\end{equation}  
The fitting parameters here are $p_{iv}=0.058$ and $\alpha_{iv}=15$.

Comparison of Figs.~\ref{fig03}a and \ref{fig04}a, indicates that
a more intensive use of ivermectin in the human population is qualitatively
equivalent to a shorter gametocytemia time needed to mantain disease
prevalence. The outcome of massive administration of ivermectin in a
fraction of the human population reveals strong correlation with an
effective reduction on the duration of positive gametocytemia.
Consequently, both the probability of ivermecting treatment
$p_{iv}$ and annual malaria incidence are anticorrelated, as shown in
Fig.~\ref{fig04}b.
Here, we run 10 simulations for each value of $p_{iv}$ ranging from
0.020 to the value 0.058, which corresponds to the obtained
critical value at the phase transition in Fig.~\ref{fig04}a.
The linear regression in Fig.~\ref{fig04}b yields
\begin{equation}
  I_{mal_{piv}} = 97 - 1509 p_{iv}
  \label{Reg02}
\end{equation}
with a Pearson correlation of $r^{2} = 0.9499$ and a $p$-value
of $P<0.001$.
Similarly as what we discussed above for Eq.~(\ref{Reg01}),
here we observe positive incidence only for
values of ivermectin $p_{iv}\lesssim 97/1509 \sim 0.064 \lesssim p_{iv}^{(t)}$.

\subsection{Combined use of gametocidal agents and ivermectin: a copula approach for predicting optimal administration intensities}
\label{subsec:combination}

\begin{figure}[t]
\centering
\includegraphics[width=0.48\textwidth]{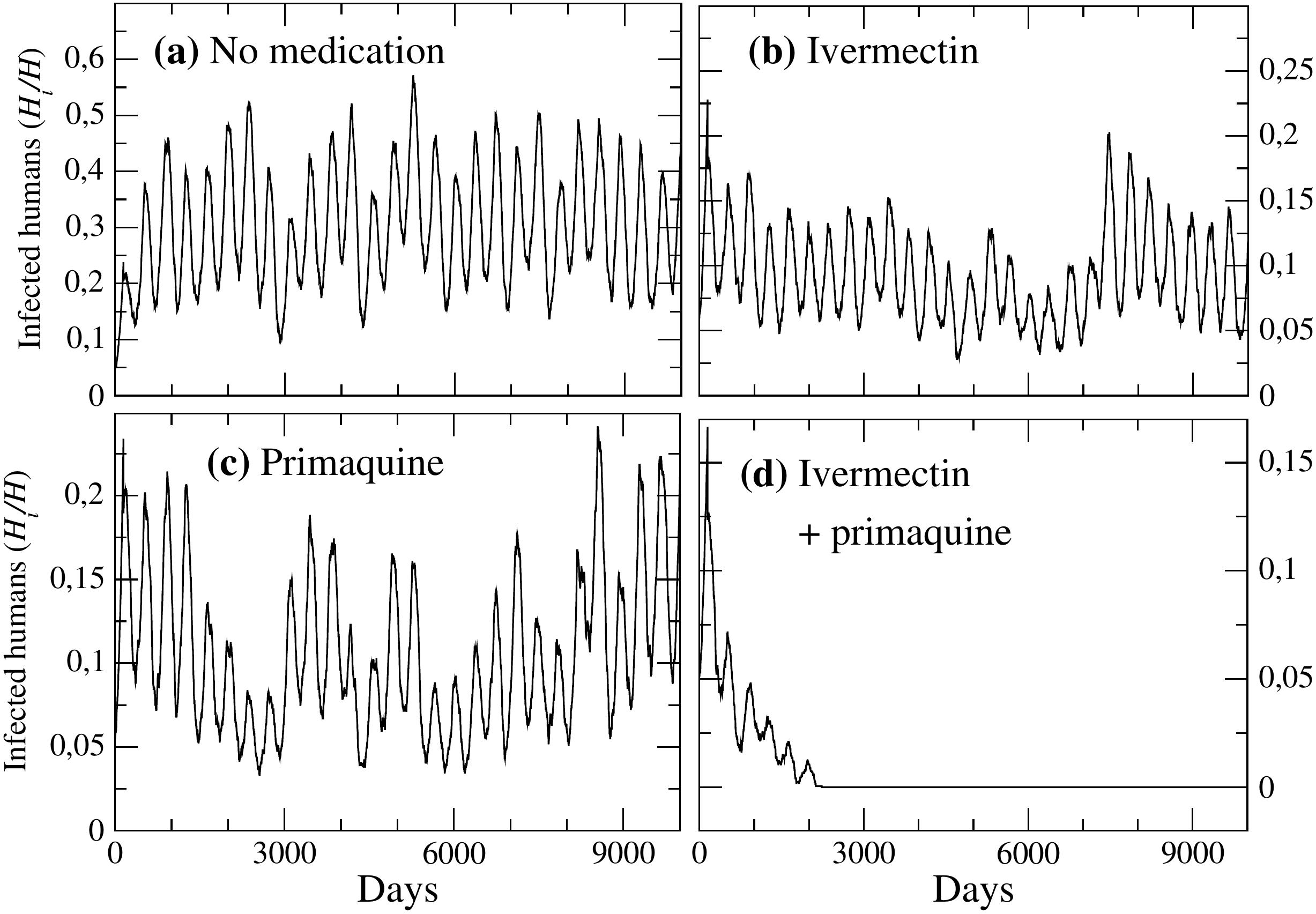}
\caption{\protect
  {\bf (a)} The evolution of the number of infected humans
  in an epidemic status with $w_{h}=0.6$ with $p_{iv}=0$.
  {\bf (b)} Application of ivermectin treatment with $p_{iv}=0.05$
  to the situation shown in (a), keeping positive gametocytemia at
  $w_{h}=0.6$.
  {\bf (c)} Application of gametocytemia reduction with primaquine
  from $w_{h}=0.6$ to $w_{h}=0.467$, without ivermectin ($p_{iv}=0$).
  {\bf (d)} Combined gametocytemia reduction with primaquine and
  ivermectin treatment in epidemic status, with $w_{h}=0.467$ with
  $p_{iv}=0.05$.}
\label{fig05}
\end{figure}

As shown in the previous Sec.~\ref{subsec:ivermectin}, in a stable epidemic
status, with 90 days of positive gametocytemia ($w_h=0.60$), after the use
of ivermectin in 5\% of the human population, there is a reduction in
the fraction of infected human hosts. Compare Fig.~\ref{fig05}a
with Fig.~\ref{fig05}b.
However, this reduction is not robust enough to achieve complete disease
eradication.
Similarly, after a reduction in the days of positive gametocytemia,
namely from 90 to 70 days, with no ivermectin treatment, there is a
weakening in disease transmission, although also not robust enough to
achieve eradication (see Fig.~\ref{fig05}c).
But combining both effects, namely with a gametocytemia reduction from 90 to 70 days and ivermectin preventive treatment in 5\% of the population, disease eradication is rapidly attained (Fig.~\ref{fig05}d).


Apparently, the combination of these separate strategies may lead to
a stronger action in suppressing malaria infection in the human population.
Our quantitative analysis however provides a framework for deriving
an estimation of how strong these strategies should be, when
used in combination, in order to achieve full disease eradication.
Assuming both factors to be independent from each other,
a first order approximation to estimate the number of infected humans
would be $\hat{H} = F_g(w_h)F_{iv}(p_{iv})H$ and eradication
would be the region in parameter space $(w_h,p_{iv})$ satisfying
\begin{equation}
  F_g(w_h)F_{iv}(p_{iv}) < \frac{1}{H} \, .
\end{equation}

The fact that the time period of the parasite development in the mosquito is
generally longer than 10 days (see Tab.~\ref{tab01}), may explain the reason
for the effectiveness of ivermectin in preventive campaigns directed to other
endemic parasites in Africa.
However, this effect may not be related to an overall reduction in the number
of mosquitoes, but also to a selective interference in the process of
parasite development towards sporozoite inside the mosquito,
and a preferential killing of infected mosquitoes.
Therefore, both factors are correlated and the prediction presented
above is biased towards a worst-case scenario.
\begin{figure*}[t]
\centering
\includegraphics[width=0.9\textwidth]{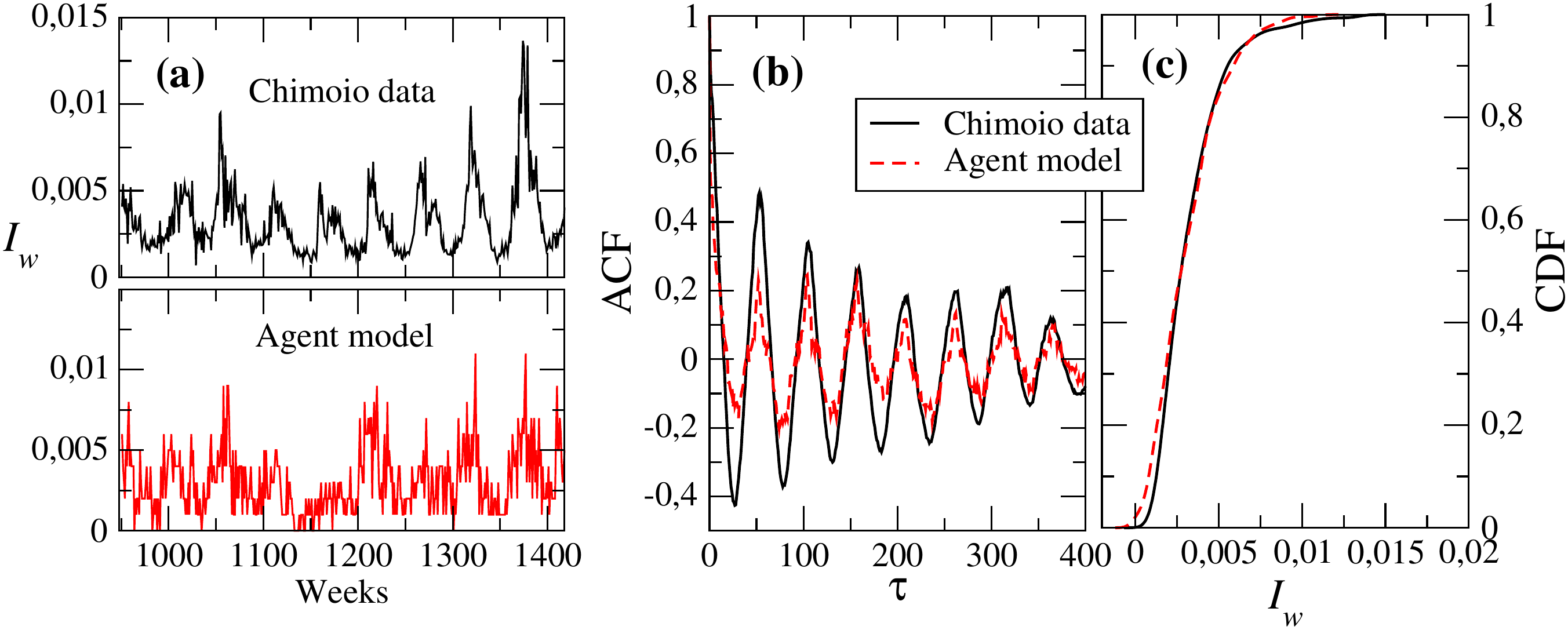}
\caption{\protect
  {\bf (a)}
  Time series of weekly malaria incidence $I_w$, comparing empirical
  data from Ref.~\cite{Ferrao2017b} (top) against data from
  one realization of the agent model (bottom),
  where 64 days of gametocytemia was used ($w_{h} = 0.427$).
  {\bf (b)} Auto-correlation function of the weekly malaria incidence
  from the empirical data (solid line),
  compared with the auto-correlation from the agent model simulation.
  {\bf (c)}
  Cumulative density functions (CDF) of the malaria incidence for both
  empirical and simulation sets of data.
  In all cases $p_{iv}=0$.}
\label{fig06}
\end{figure*}

\section{Model validation and consistency tests:
  comparison with malaria transmition in Chimoio}
\label{sec:validation}

In this section, we validate the agent model simulations against empirical
data, namely time series at weekly intervals collected at Chimoio region
in Mozambique \cite{Ferrao2017b}.
In this African region, malaria is endemic revealing a trend, which increases
during the four to five months of the wet season (high transmission season)
and decreases during the rest of the year (low transmission season).
The empirical time series includes a total of 490 561 malaria cases
in a population of $H=324 816$ human individuals,
recorded from January 1st 2006 to December 31st 2014.
During these 9 years, weekly malaria incidence $I_w$ was analyzed in both
the empirical set of data and in simulated data generated by the agent
model.
Figure \ref{fig06}a shows both time series during 9 years (468 weeks).
From the total time of 30 years covered by the simulation, we discarded
the first three years (156 weeks) of the simulation, in order to weaken
the influence of the initial conditions.
From the remaining 27 years we extracted the 9 years interval, that best
correlate with the empirical series.
The maximum of the correlation is achieved from week 951 to week 1418,
as in the horizontal axis label of Fig.~\ref{fig06}a.
The simulation uses 64 days of positive gametocytemia ($w_{h}=0.427$),
which relates to a marginally higher
human-to-mosquito disease transmission efficiency, than that found at
phase transition (cf.~Fig.~\ref{fig03}a).
Ivermectin was not considered ($p_{iv}=0$) at this stage.

In Fig.~\ref{fig06}b we plot the auto-correlation functions of the
empirical data and of the simulation.
The autocorrelation is defined as
\begin{equation}
  \gamma(\tau) = \frac{\langle (I_w(t+\tau) - \bar{I}_w)(I_w(t) - \bar{I}_w)\rangle}{\sigma_{I_w}^2} \, ,
  \label{acf}
\end{equation}
where $\bar{I}_w$ and $\sigma_{I_w}^2$ are respectively the mean and variance
of the incidence series and $\langle\cdot\rangle$ represents the average over
time $t$.
Apart from a deviation of the
local extremes, the periodicity of the simulated scenario matches
rather well with the real seasonal oscillation period ($\sim 1$ year).
To quantify the similarity between real data and agent model simulation
with computed the usual performance metrics, namely
the mean absolute error (MAE)
\begin{equation}
  \mathrm{MAE} = \frac{1}{n} \sum_{t=1}^n \left|
  \hat{I}_w(t) - I_w(t) \right| \, 
  \label{eq:MAE}
\end{equation}
with $\hat{I}_w(t)$ and $I_w(t)$ representing the simulated and
real incidence value and $n=468$ (weeks),
the mean absolute percentual error (MAPE)
\begin{equation}
  \mathrm{MAPE} = \frac{1}{n} \sum_{t=1}^n \left|
  \frac{\hat{I}_w(t) - I_w(t)}{I_w(t)}\right| \, 
  \label{eq:MAPE}
\end{equation}
and the root mean square error
\begin{equation} 
  \mathit{RMSE} = \left [ \frac{1}{n} \sum_{t=1}^n (\hat{I}_w(t) - I_w(t))^2 \right ]^{1/2} \, .
  \label{eq:MSE}
\end{equation}
The computation yields MAE$=0.00152$, MAPE$=0.558$ and 
RMSE$=9.54\times 10^{-5}$.
The simulation are within fluctuations of $50\%$ of real incidence values.

We also compare the distribution of simulated and real incidence values,
as plotted in Fig.~\ref{fig06}c: the cummulative
distributions match rather well, with a small Kolmogorov-Smirnov
score (0.22) having a $p$-value smaller than $0.001$.

Importantly, in all simulations, mosquito and human infection is strongly
related, showing a similar oscillatory pattern.
Moreover, only a small fraction of the mosquito population survived beyond
the parasite development in the mosquito (10 days), which leads to
a strong correlation between endemic prevalence in humans and mosquitoes
in all endemic scenarios \cite{McKenzie2005}.
\begin{table}[t]
\begin{center}
\begin{tabular}{lcc}
\hline
\textsc{Description} &\textsc{Theory} & \textsc{Model}\\ 
\hline
\hline
{Basic reproductive number} ($R_0$) & {\bf 1.619} & {\bf 0.973}\\ 
{Annual entom.~inoculation rate} ($EIR$) & {\bf 0.961} & {\bf 0.965}\\
\hline
\hline
Fraction of infected mosquitoes ($m$) & 2 & 2\\ 
Human feeding rate ($a$) & 0.238 & 0.239 \\ 
Sporozoite rate ($Z$) & --- & 0.006 \\
Force of infection ($\lambda$)& 0.47 & 0.284 \\
Mosquito-to-Human transmission ($b$) & 0.18 & 0.108\\ 
Human-to-Mosquito transmission ($c$) & 0.091 & 0.090\\ 
Human recovery rate ($q_{h}$) & 0.011 & 0.011\\
Mosquito daily mortality ($q_{m}$) & 0.1 & 0.1\\ 
\hline
\end{tabular}
\end{center}
\caption{\protect
  Classical Ross parameters from theory and model simulation for
  endemic scenario A, close to phase transition, i.e.~with
  68 days of gametocytemia ($w_{h}= 0.453$). The two main quantities,
  reproductive number $R_0$ and the yearly entomological inoculation
  ratio $EIR$, are within the classical theoretical values.}
\label{tab03}
\end{table}

Our model assumes rules based on classical and neoclassical assumptions,
including several quantities from the classical Ross-MacDonald
model \cite{Ross1915,Macdonald1957,DLSmith2004}.
The two main quantities in this classical model are
the annual entomological inoculation ratio $EIR$, which is defined
as the number of bites per year on a human host from an infectious mosquito,
and the reproductive number $R_0$, which represents the number of
infected humans generated from one single infectious mosquito in a
population of susceptible and non-immune individuals.

For evaluation of our model, we compare the values obtained in our
simulations with the expected theoretical ones, which are
given in Refs.~\cite{Ross1915,DLSmith2004}.
Results are given in Tab.~\ref{tab03}.
For estimating the annual entomological inoculation ratio we use the
definition
\begin{equation}
  EIR = 365\, m \, a \, Z ,
\end{equation}
where $m$ is the mosquito density (number of mosquitoes per human individual,
$m = \frac{N_m}{N_h}$),
$a$ is the human feeding rate given by (see Tab.~\ref{tab01} and
Sec.~\ref{sec:model}), 
$a = (p_Q n_b (1-u) \sigma)/\tau_s$,
and $Z$ is the sporozoite rate (fraction of infectious mosquitoes).
For estimating the reproductive number, we use the Ross definition
\cite{Ross1915}
\begin{equation}
  R_{0} = \frac{ma^{2}bc}{q_{h}q_{m}} \, ,
  \label{R0-Ross}
\end{equation}
where $b$ is the mosquito-to-human transmission efficiency,
$b = k_h (1-\bar{v})$,
$c$ is the human-to-mosquito transmission efficiency,
$c = k_m w_h$, and
$q_h$ and $q_m$ are given by Eqs.~(\ref{qh}) and (\ref{qm}) respectively.
As indicated in Tab.~\ref{tab03}, the expected theoretical values
of both these quantities are well reproduced by the simulations.
The lower value of $R_0$ in the simulation, when compared with Ross theory
is due to the fact that the $b$ value in the simulation only takes
into account bites from infectious mosquitoes.

\section{Discussion and conclusion: towards medical strategies}
\label{sec:discussion}

We introduce an agent model for assessing the effect of gametocytemia
and drug administration in epidemiological scenarios of malaria.
Our model was calibrated by considering various aspects of the disease
dynamics and supported by field data.
We uncover the existence of a phase transition between an absortion
state with disease eradication and an endemic/epidemic regime.

Because several parameters from our model were based on the Mozambique epidemic environment, validation of the model implementation took place by the comparison with field collected data series for malaria incidence in the typical seasonal endemic malaria region of Chimoio, Mozambique. Importantly, this field data time-series covered a long time period of malaria incidence, namely 9 years \cite{Ferrao2017b}.
Although the parameter values in Tab.~\ref{tab01} are case-dependent, they are within the range of typical values described in the literature.

In complex models, phase transition stands as a critical concept of stochastic simulation. Its precise definition is useful to identify the occurrence of state transition between disease eradication and endemic stability, which can be used for better preventive planning. At critical equilibrium points, malaria transmission dynamics was defined taking into account the predicted rational use of anti-malarial strategies in the near future.
 
Special attention was given to the role of gametocytemia in human-to-mosquito transmission. All our model simulations assumed a duration of positive gametocytemia in the range of 0.387 to 0.733  of total infection time. With a small variation in gametocytemia prevalence it was possible to define all tested transition phases. These small changes in gametocytemia were considered as a model for effective gametocidal treatment, such as the administration of primaquine or methylene blue \cite{ Karl2011, Kuehn2010, Eziefula2012, Sutanto2013, John2016, Goncalves2016, Lin2017}. Transition phases were clearly defined, promoting a better understanding of the disease dynamics, as well as of the points of sudden stochastic transition from epidemic prevalence to disease eradication.

In the present model, we also analyzed  preventive intervention with ivermectin, a well known agent with capability of interfering with human-to-mosquito transmission. An intervention with ivermectin may be highly selective in targeting recently infected mosquitoes, killing the mosquito before the complete development of the parasite in the mosquito. This aspect bears no relation to gametocytemia prevalence \cite{Chaccour2010, Kobylinski2012, Ouedraogo2015}. Apparently, the role of mosquito mortality from ivermectin in disease transmission does not significantly overlap with the effect of gametocidal drug intervention. 

The detailed biochemical mechanisms that trigger gametocytogenesis in
{\it Plasmodium} are not well known.
However, it is known that this process may be influenced by host immunity and anti-malaria therapy \cite{Karl2011,Kuehn2010}.
For human-to-mosquito transmission to be effective, male and female stage V gametocytes must be present in the blood during mosquito feeding. 
Once inside mosquito midgut, gametocytes will mature to gametes promoting fertilization and maturation to zygote stage, ookinete, oocyst, and finally to the sporozoite, the infectious form of the parasite present in mosquito salivary glands.
Common gametocidal drug agents (primaquine, artemisinin and methylene blue \cite{Eziefula2012, Sutanto2013, John2016, Goncalves2016,  Lin2017, Peatey2009, Bosson-Vanga2018}) usually fail to act in early stages of gametocyte maturation. But their inhibitory action on Gametocytes in stage V may be very effective in reducing the time of gametocytemia duration \cite{Kuehn2010}.

Vector control, by itself, is not enough to eradicate disease transmission. Long standing cyclic positive gametocytemia in a few human individuals  may perpetuate transmission for a long time and more attention should be directed towards human disease reservoirs as possible hot-spots for chronic mosquito infection. Preventing mosquito infection from these hot-spot human reservoirs by reducing the time of positive gametocytemia with the help of a selective mosquito-killing-after-bite  preventive drug strategy with ivermectin, may turn out to be a more effective strategy in the fight against malaria.

The combined intervention may also be useful in reducing pressure
in areas where drug resistance is becoming major
problem \cite{Dondorp2010, Dama2017}. These results seem to indicate that
such a theoretical possibility may deserve serious consideration in future
malaria prevention campaigns. 

Dynamical aspects of human therapy with drug agents such as artemisinin or quinine (with specific intervention in disease status and gametocytemia probability), population heterogeneity  and human migration were not included in the present analysis. Model simulations assumed the existence of a typical isolated African village with limited drug therapy availability.

Our computational model allowed us to test the combined use of different preventive interventions with antimalarial agents like ivermectin (killing mosquitoes during parasite's development) or primaquine (gametocytemia reduction) that could significantly influence disease outcome, and therefore contribute to a better knowledge of disease transmission dynamics in different endemic scenarios. With the present model, it is possible to recreate simulations for different disease regions with specific seasonality conditions, and to anticipate events as a result of selective interventions in certain human subgroups in all simulations.

From the main findings of this work, a set of valuable insights are 
possible.
First, in endemic locations small differences in gametocytemia prevalence in human populations, obtained from preventive intervention in a small fraction of the population with gametocidal drugs \cite{Eziefula2012, Sutanto2013, John2016, Goncalves2016,  Lin2017, Peatey2009}, may result in very different outcomes, despite the relative stability of classical human-to-mosquito infectiousness parameter \textit{c}.

Second, the demonstrated mosquitocidal properties of ivermectin in the first days after a mosquito feed, may potentiate the effect of gametocidal agents with drastic interference in human-to-mosquito transmission efficiency. This preventive action may also benefit from its combined use with LLIN/ITN/IRS.

Third, our model indicates that with a combined ivermectin and primaquine scissor-like intervention, malaria eradication may be possible in a small African village after a short time period.

\section*{Acknowledgments}

We thank Miguel Prud\^encio  (IMM - Lisbon) for his help in defining model parameters, and for his crucial remarks in discussing the model design. 

\section*{References}
\bibliographystyle{elsarticle-num}
\bibliography{malaria_bibliography.bib}

\end{document}